\documentclass[aps]{revtex4}
\usepackage{eurosym}
\usepackage{amsfonts}
\usepackage{amsmath}
\usepackage{amssymb,epsf}
\usepackage{color}
\usepackage{xcolor}
\usepackage{graphicx}
\usepackage{epstopdf}
\usepackage{float}
\usepackage{caption}
\usepackage{subfig}
\usepackage{hyperref}

\begin{document}

\title{Super-entropy bumblebee AdS black holes}
\author{B. Eslam Panah$^{1,2,3}$\footnote{%
email address: eslampanah@umz.ac.ir}}
\affiliation{$^{1}$ Department of Theoretical Physics, Faculty of Basic Sciences,
University of Mazandaran, P. O. Box 47416-95447, Babolsar, Iran\\
$^{2}$ ICRANet-Mazandaran, University of Mazandaran, P. O. Box 47416-95447,
Babolsar, Iran\\
$^{3}$ ICRANet, Piazza della Repubblica 10, I-65122 Pescara, Italy}

\begin{abstract}
Motivated by the effect of the bumblebee field on thermodynamic instability
in (non)extended phase space, we study the thermodynamic instability for the
bumblebee AdS black holes. For this purpose, first, we evaluate the effect
of the bumblebee field (or Lorentz-violating parameter) on the event horizon
for AdS black holes. Then, in non-extended phase space, we study the effect
of the bumblebee field on the heat capacity and the Helmholtz free energy to
investigate the local and global thermal stability areas, respectively.
Next, we extend our study on the extended phase space by seeking on stable
area by using the heat capacity at constant pressure ($C_{P}$). Finally, we
evaluate the super-entropy black hole condition and indicate that the
bumblebee AdS black holes are super-entropy black holes when $l>0$, which is
consistent with the condition $C_{P}<0$.
\end{abstract}

\maketitle

\section{Introduction}

Bumblebee gravity is an alternative to general relativity (GR) as a
mechanism for Lorentz symmetry violation. This theory includes the bumblebee
field as a vector field in the Einstein-Hilbert action. The bumblebee field (%
$B_{\mu }$) is quadratically coupled to the Ricci tensor ($R_{\mu \nu }$),
and it can adopt a nontrivial background configuration through a
cosmological potential term $V$. The action of bumblebee gravity in the
presence of the cosmological constant is given by \cite{Maluf2021} 
\begin{equation}
\mathcal{I}=\int_{\partial \mathcal{M}}d^{4}x\sqrt{-g}\left[ \frac{%
R-2\Lambda }{2\kappa }+\frac{\xi \left( B^{\mu }B^{\nu }R_{\mu \nu }\right) 
}{2\kappa }-\frac{B_{\mu \nu }B^{\mu \nu }}{4}-V\left( B^{\mu }B_{\mu }\pm
b^{2}\right) +\mathcal{L}_{matter}\right] ,  \label{action}
\end{equation}%
where $\kappa =\frac{8\pi G}{c^{4}}$, and $\Lambda $ is the cosmological
constant. In addition, $\xi $ plays the role of a coupling constant that
accounts for the nonminimum interaction between the bumblebee field and the
Ricci tensor \cite{Bailey2006, Malufetal2013}. Moreover, $B_{\mu \nu
}=\partial _{\mu }B_{\nu }-\partial _{\nu }B_{\mu }$ represents the
bumblebee field strength, and $\mathcal{L}_{matter}$ describes the matter
and additional couplings with the field $B_{\mu }$.

The bumblebee gravity theory involves the spontaneous violation of Lorentz
symmetry. Lorentz-violating effects can be observed in various contexts,
including string theory \cite%
{KosteleckyS1989a,KosteleckyS1989b,KosteleckyS1989c,KosteleckyP1991}, loop
quantum gravity \cite{GambiniP1999,Ellis2000}, noncommutative spacetime \cite%
{Carroll2001,Mocioiu2000,Ferrari2007}, Horava-Lifshitz gravity \cite%
{Horava2009}, warped brane worlds \cite{Rizzo2010,Santos2013}, and gravity's
rainbow \cite{MagueijoS2004}. A framework called the Standard Model
Extension (SME), proposed by Colladay and Kostelecky \cite%
{Colladay1997,Colladay1998}, provides a suitable way to account for
Lorentz-violating effects on the behavior of elementary particles. This
framework is based on the idea of spontaneous Lorentz symmetry breaking in
string theory \cite{KosteleckyS1989a}.

SME is an effective field theory that includes additional gauge invariant
terms. These additional terms are obtained by contracting the physical
standard model fields with fixed background tensors \cite%
{Bluhm2005,Bluhm2008}. Some research has been carried out on various sectors
of the SME, leading to determining constraints on the Lorentz-violating
parameters. For example, investigations into CPT symmetry violation \cite%
{Bluhm1997,Bluhm1999,Bluhm2000,Bluhm2002}, the gauge CPT-odd/even sectors 
\cite{Carroll1990,Andrianov1998,Lehnert2004,Kaufhold2006,Belich2013}, the
fermion sector \cite{Colladay2001,Lehnert2004b,Altschul2004,Shore2005},
radiative corrections \cite%
{Jackiw1999,Chung2001,Battistel2001,Scarpelli2001,Mariz2005,Nascimento2007,Scarpelli2008,Cima2010}%
, and photon-fermion interactions \cite%
{Gazzola2012,Baeta2012,Agostini2012,Brito2013}.

Traversable wormholes in bumblebee gravity were discussed in Ref. \cite%
{Ovgun2019}, which satisfied the energy conditions for normal matter and the
flare-out conditions near the throat. Subsequently, the effect of bumblebee
gravity on the gravitational lensing of the wormhole spacetime was studied,
demonstrating that the bumblebee field effect resulted in a non-trivial
global topology of the wormhole spacetime. In addition, the effect of the
bumblebee field on the quasinormal mode frequencies of traversable wormholes
has been studied using the WKB approximation method for both scalar and
gravitational perturbations \cite{Oliveira2019}.

In the context of black holes, researchers examined how Lorentz symmetry
breaking affects the Hawking radiation of a Schwarzschild-like black hole
within bumblebee gravity. They found that, regardless of the spin of the
emitted particle, the generalized uncertainty principle alters the Hawking
temperature of the black hole \cite{Kanzi2019}. In Refs. \cite%
{Kanzi2021,Kanzi2022}, the effect of the bumblebee field on the quasinormal
modes and greybody factors of Kerr-like black holes has been investigated in
detail. In addition, the influence of bumblebee gravity parameters on the
behavior of test particles orbiting a slowly rotating, axially symmetric,
charged black hole has been examined in Ref. \cite{Mustafa2025}. The impact
of the bumblebee field on the shadows of black holes has been examined in
reference to EHT observations in Refs. \cite{Shadow1,Shadow2}. Other
interesting properties of black holes in the context of bumblebee gravity
have been examined in Refs. \cite{BHBB1,BHBB2,BHBB3,BHBB4,BHBB5}.

The effect of bumblebee fields in the context of cosmology has been studied
in the literature. For example, the dynamic equations governing the
evolution of the Universe, along with their properties and physical
significance within the framework of bumblebee gravity, are examined. The
results indicate that a late-time de Sitter expansion of the Universe can be
replicated, and efforts are made to constrain the parameters of the
potential that drives spontaneous symmetry breaking \cite{Capelo2015}. In
Ref. \cite{MalufJCAP2021}, a bumblebee field was employed to generate
cosmological anisotropies. This bumblebee field serves as a source of
anisotropies, resulting in the emergence of a preferred axis. Consequently,
a fraction of the cosmic anisotropies can be attributed to the violation of
Lorentz symmetry. Additionally, an upper bound on the bumblebee field has
been determined using the quadrupole and octopole moments of the cosmic
microwave background radiation. Moreover, it was demonstrated that both
anisotropy and the Bumblebee field significantly influence cosmic evolution,
leading to a prolonged matter-dominated phase compared to standard
cosmology. This suggests that these factors may offer a deeper understanding
of the universe's structure and expansion dynamics \cite{Sarmah2024}. The
impact of the bumblebee field on various aspects of cosmology has been
examined in Refs. \cite{CosBB1,CosBB2,CosBB3}.

\section{Field Equations and Black Hole Solutions}

By varying the action (\ref{action}) with respect to the metric tensor $%
g_{\mu \nu }$, while keeping the bumblebee field $B_{\mu }$ fixed, the
gravitational field equations in the bumblebee gravity is obtained \cite%
{Maluf2021} 
\begin{equation}
G_{\mu \nu }+\Lambda g_{\mu \nu }=\kappa \left( T_{\mu \nu }^{B}+T_{\mu \nu
}^{M}\right) ,  \label{FE1}
\end{equation}%
where $G_{\mu \nu }$ is the Einstein's tensor. $T_{\mu \nu }^{B}$ is the
energy-momentum tensor of the bumblebee field in the following form \cite%
{Maluf2021} 
\begin{eqnarray}
T_{\mu \nu }^{B} &=&2V^{^{\prime }}B_{\mu }B_{\nu }+B_{\mu }^{~\ \alpha
}B_{\nu \alpha }-\left( V+\frac{1}{4}B_{\alpha \beta }B^{\alpha \beta
}\right) g_{\mu \nu }+\frac{\xi }{2\kappa }\left[ B^{\alpha }B^{\beta
}R_{\alpha \beta }g_{\mu \nu }-2B_{\mu }B^{\alpha }R_{\alpha \nu }-2B_{\nu
}B^{\alpha }R_{\alpha \mu }\right.  \notag \\
&&  \notag \\
&&\left. +\nabla _{\alpha }\nabla _{\mu }\left( B^{\alpha }B_{\nu }\right)
+\nabla _{\alpha }\nabla _{\nu }\left( B^{\alpha }B_{\mu }\right) -\nabla
^{2}\left( B_{\mu }B_{\nu }\right) -g_{\mu \nu }\nabla _{\alpha }\nabla
_{\beta }\left( B^{\alpha }B^{\beta }\right) \right] ,  \label{FE2}
\end{eqnarray}%
in which the operator ${^{\prime }}$ means derivative with respect to the
potential argument. Also, $T_{\mu \nu }^{M}$ is the energy-momentum tensor
of the matter field. In the above action, $g=det(g_{\mu \nu })$ is the
determinant of metric tensor $g_{\mu \nu }$. Hereafter, we consider $G=c=1$.

Furthermore, there is another equation of motion for $B^{\mu }$ which is
getting by varying the action (\ref{action}) with respect to the bumblebee
field 
\begin{equation}
\nabla _{\mu }B^{\mu \nu }=2\left( V^{^{\prime }}B^{\nu }-\frac{\xi }{%
2\kappa }B_{\mu }R^{\mu \nu }\right) .  \label{FE3}
\end{equation}

We consider a four-dimensional static spacetime as 
\begin{equation}
ds^{2}=-f(r)dt^{2}+\left( 1+l\right) \frac{dr^{2}}{f(r)}+r^{2}\left( d\theta
^{2}+\sin ^{2}\theta d\varphi ^{2}\right) ,  \label{Metric}
\end{equation}%
where $f(r)$ is the metric function. Also, the bumblebee field or the
Lorentz-violating parameter is represented in that metric by $l=\xi b^{2}$.
It is notable that, $b=\sqrt{b_{\mu }b^{\mu }}$, and $b_{\mu }$ is defined
in the following form 
\begin{equation}
B_{\mu }=b_{\mu }=\left( 0,b_{r}\left( r\right) ,0,0\right) ,  \label{FE4}
\end{equation}%
which $b_{\mu }$ is the background field, and it is a spacelike vector
purely radial. Also, to keep the signature of the mentioned spacetime (\ref%
{Metric}), we have to impose the constraint $l>-1$.

Using Eqs. (\ref{FE1})-(\ref{FE4}), the metric function is given by \cite%
{Maluf2021} 
\begin{equation}
f(r)=1-\frac{2m_{0}}{r}-\left( 1+l\right) \frac{\Lambda }{3}r^{2}.
\label{f(r)}
\end{equation}%
where $\Lambda =\frac{\kappa \lambda }{\xi }$, it is known an effective
cosmological constant. Also, $\lambda $ is a Lagrange-multiplier field (see
Refs. \cite{Bluhm2008,Maluf2021}, for more details).

Considering the spacetime in Eq. (\ref{Metric}), and the metric function (%
\ref{f(r)}), we can obtain the Ricci and Kretschmann scalars in the
following forms 
\begin{eqnarray}
R &=&4\Lambda +\frac{2l}{\left( 1+l\right) r^{2}}, \\
&&  \notag \\
R_{\alpha \beta \gamma \delta }R^{\alpha \beta \gamma \delta } &=&\frac{%
8\Lambda ^{2}}{3}+\frac{8\Lambda l}{3\left( 1+l\right) r^{2}}+\frac{4l^{2}}{%
\left( 1+l\right) ^{2}r^{4}}+\frac{16m_{0}l}{\left( 1+l\right) ^{2}r^{5}}+%
\frac{48m_{0}^{2}}{\left( 1+l\right) ^{2}r^{6}},
\end{eqnarray}%
which indicates a curvature singularity located at $r=0$, because the Ricci
and Kretschmann scalars diverge at $r=0$. Also, it is finite for $r\neq 0$.
On the other hand, the asymptotical behavior of the Ricci and Kretschmann
scalar are given by $\lim_{r\longrightarrow \infty }R\longrightarrow
4\Lambda $, and $\lim_{r\longrightarrow \infty }R_{\alpha \beta \gamma
\delta }R^{\alpha \beta \gamma \delta }\longrightarrow \frac{8\Lambda ^{2}}{3%
}$, which shows the spacetime will be asymptotically (A)dS, when the
cosmological constant is $\Lambda >0$ ($\Lambda <0$).

\begin{figure}[tbh]
\centering
\includegraphics[width=0.4\textwidth]{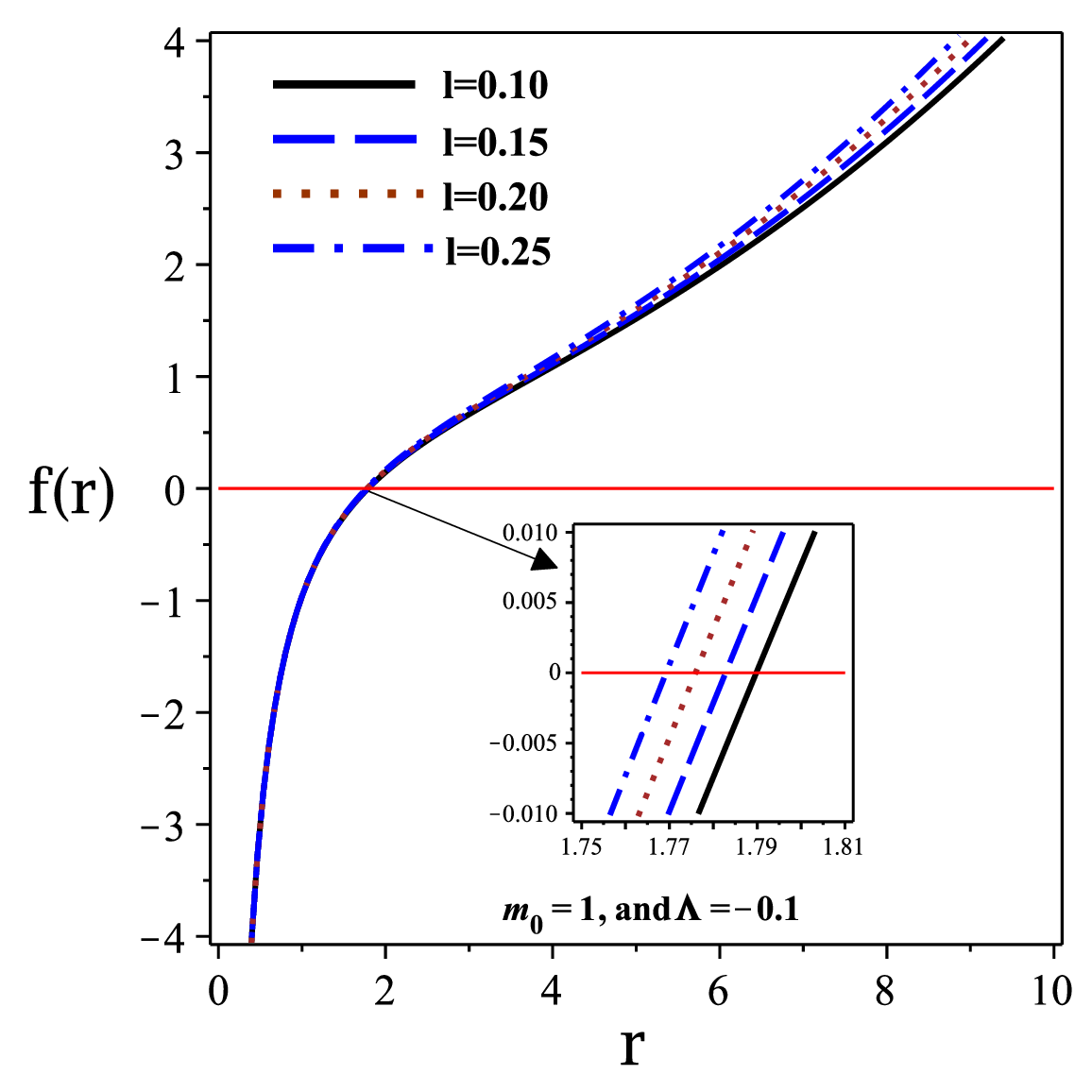} \newline
\includegraphics[width=0.4\textwidth]{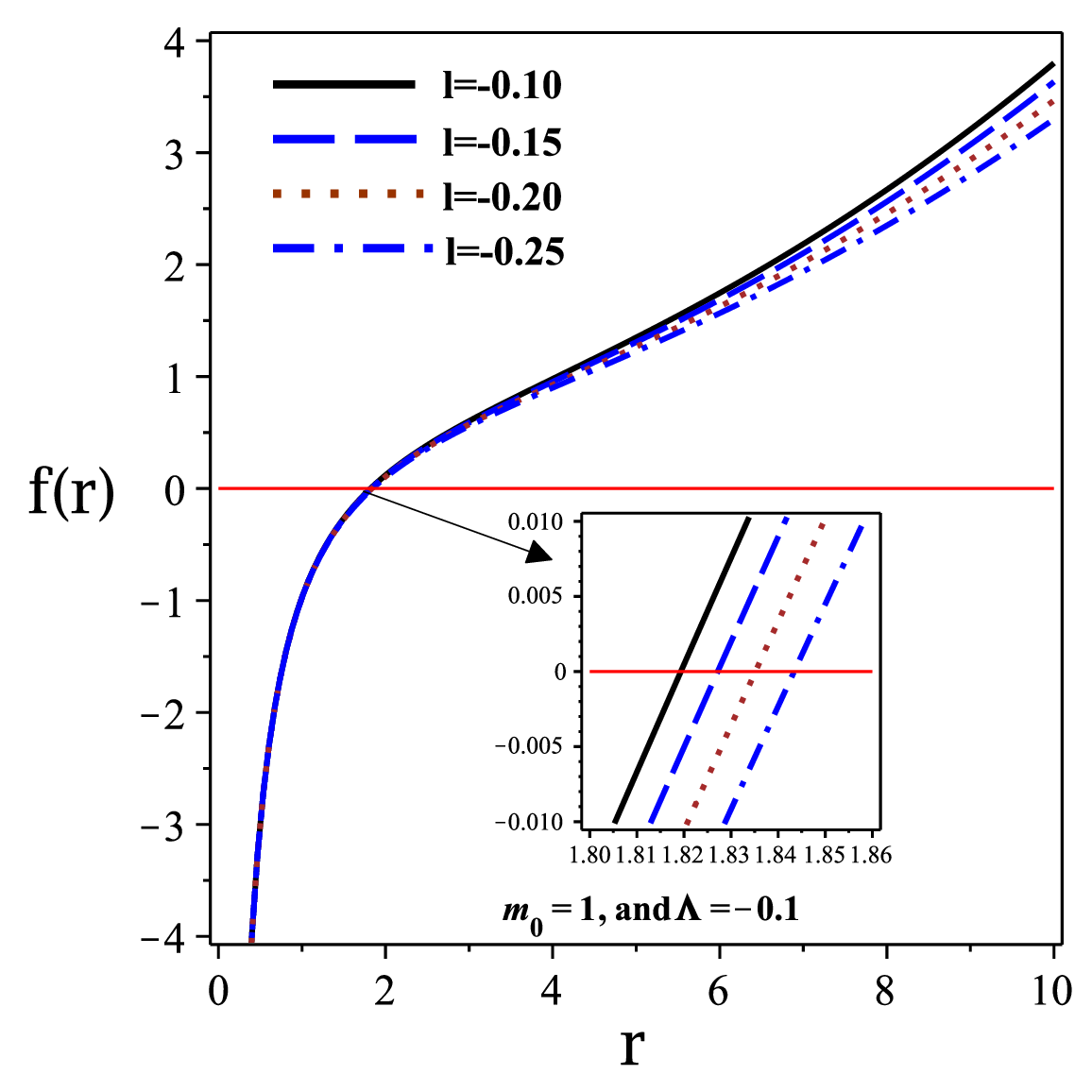} \newline
\caption{$f(r)$ versus $r$ for different values of parameters.}
\label{Fig1}
\end{figure}

To study the effects of various parameters on the event horizon of the black
hole, we plot the metric function $f(r)$ versus $r$, in Fig. \ref{Fig1}. Our
findings indicate that for the positive Lorentz-violating parameter ($l>0$),
the radius of the event horizon decreases by increasing $l$ (see the up
panel in Fig. \ref{Fig1}). For the negative Lorentz-violating parameter ($%
l<0 $), we encounter with large black holes when $|l|$ increases (see the
down panel in Fig. \ref{Fig1}). 

\section{\textbf{Thermodynamics}}

In order to investigate the thermodynamic properties of the bumblebee black
hole solutions, it is crucial to establish the relationship between the mass
($m_{0}$), the event horizon radius ($r_{+}$), and the cosmological
constant. This can be achieved by setting $g_{tt}=f(r)$ equal to zero (i.e. $%
g_{tt}=f(r)=0$). So, we have 
\begin{equation}
m_{0}=\frac{\left( 3-\left( 1+l\right) \Lambda r_{+}^{2}\right) r_{+}}{6}.
\label{mm}
\end{equation}

We calculate the surface gravity of the bumblebee black holes in order to
determine the Hawking temperature of these black holes, which is 
\begin{equation}
\kappa =\left. \frac{g_{tt}^{\prime }}{2\sqrt{-g_{tt}g_{rr}}}=\right\vert
_{r=r_{+}}=\left. \frac{f^{\prime }(r)}{2\sqrt{1+l}}\right\vert _{r=r_{+}}.
\label{k}
\end{equation}%
Considering the metric function (\ref{f(r)}), and substituting the mass (\ref%
{mm}) into equation (\ref{k}), one can calculate the surface gravity as $%
\kappa =\frac{1}{2\sqrt{1+l}r_{+}}-\frac{\left( 1+l\right) \Lambda r_{+}}{2%
\sqrt{1+l}}$. By using the Hawking temperature, $T=\frac{\kappa }{2\pi}$, we
can express it in the following form 
\begin{equation}
T=\frac{1}{4\pi \sqrt{1+l}r_{+}}-\frac{\left( 1+l\right) \Lambda r_{+}}{4\pi 
\sqrt{1+l}}.  \label{TemF(R)CPMI}
\end{equation}
It is clear that there is a divergence point at $l=-1$ for the temperature.
To avoid this divergence, we cannot consider $l=-1$. On the other hand, we
encounter imaginary values for $l<-1$. Therefore, the permit values of $l$
is in the range $l>-1$.

To calculate the entropy of black holes in bumblebee gravity, we can use the
area law, which states that $S=\frac{A}{4}$. In this law, $A$ represents the
horizon area and is defined as $A=\left. \int_{0}^{2\pi }\int_{0}^{\pi }%
\sqrt{g_{\theta \theta }g_{\varphi \varphi }}\right\vert _{r=r_{+}}=\left.
r^{2}\right\vert _{r=r_{+}}=4\pi r_{+}^{2}$. Therefore, we can express the
entropy of bumblebee AdS black holes as 
\begin{equation}
S=\pi r_{+}^{2}.  \label{S}
\end{equation}

By employing the Ashtekar-Magnon-Das (AMD) approach \cite{AMDI, AMDII}, the
total mass of the black hole can be computed. The formula for the total mass
is as follows 
\begin{equation}
M=\frac{m_{0}}{\sqrt{1+l}},  \label{AMDMass}
\end{equation}%
where by substituting the mass (\ref{mm}) within the equation (\ref{AMDMass}%
), yields 
\begin{equation}
M=\frac{\left( 3-\left( 1+l\right) \Lambda r_{+}^{2}\right) r_{+}}{6\sqrt{1+l%
}}.  \label{MM}
\end{equation}
Similar to the Hawking temperature, we must consider the limit $l>-1$ to
avoid of a singularity at $l=-1$, and imaginary value. So, we must respect
to limit $l>-1$.

It is easy to demonstrate that the conserved quantities and thermodynamic
quantities adhere to the first law of thermodynamics in the following manner 
\begin{equation}
dM=TdS,
\end{equation}%
where $T=\left( \frac{\partial M}{\partial S}\right) =\frac{\left( \frac{%
\partial M}{\partial r_{+}}\right) }{\left( \frac{\partial S}{\partial r_{=}}%
\right) }$, and is in agreement with Eq. (\ref{TemF(R)CPMI}).

\subsection{\textbf{Thermal stability in non-extended space}}

When examining the black hole as a thermodynamic system, our goal is to
study its local and global stability. In particular, we will examine how the
presence of the bumblebee field or the Lorentz-violating parameter ($l$)
impacts the stability of the AdS black holes at both the local and global
levels.

\subsubsection{Local stability}

Our objective in this study is to examine the local stability of AdS black
holes in bumblebee gravity. To achieve this, we will analyze the heat
capacity of these black holes. The heat capacity, in the canonical ensemble,
provides important information about the thermal structure of black holes.
It indicates whether the system is thermally stable or not, with a positive
sign indicating stability and a negative sign indicating instability.
Therefore, our goal is to calculate the heat capacity of the bumblebee black
holes and use this value to investigate their local stability.

Before extracting the heat capacity, let us first re-write the total mass of
the black hole (\ref{MM}) in terms of the entropy (\ref{S}) in the following
form 
\begin{equation}
M\left( S\right) =\frac{\left( 3\pi -\left( 1+l\right) \Lambda S\right) 
\sqrt{S}}{6\pi ^{3/2}\sqrt{1+l}},  \label{MSQ}
\end{equation}%
using the equation (\ref{MSQ}), we re-write the temperature in the following
form 
\begin{equation}
T=\frac{dM\left( S\right) }{dS}=\frac{1-\frac{\left( 1+l\right) \Lambda S}{%
\pi }}{4\sqrt{1+l}\sqrt{\pi S}}.  \label{TM}
\end{equation}

Considering Eqs. (\ref{MSQ}) and (\ref{TM}), we can obtain the heat capacity
in form 
\begin{equation}
C=\frac{T}{\left( \frac{dT}{dS}\right) }=\frac{\left( \frac{dM\left(
S\right) }{dS}\right) }{\left( \frac{d^{2}M\left( S\right) }{\partial S^{2}}%
\right) }=\frac{2\left( \frac{\left( 1+l\right) \Lambda S}{\pi }-1\right) S}{%
1+\frac{\left( 1+l\right) \Lambda S}{\pi }}.  \label{Heat1}
\end{equation}

In the context of black holes, there is an argument that the root of the
heat capacity ($C=T=0$) serves as a boundary between physical ($T>0$) and
non-physical ($T<0$) black holes. This boundary is referred to as a physical
limitation point \cite{EslamPanah2018}. In other words, the heat capacity
changes sign at this physical limit point (i.e. $T=dM(S)/dS=0$).
Additionally, it is assumed that the divergences of the heat capacity
represent the critical points of phase transition (i.e. $d^{2}M(S)/dS^{2}=0$%
) for black holes.

Using Eq. (\ref{TM}) and solving it in terms of the entropy, we get the
physical limitation point as 
\begin{equation}
S_{root}=\frac{\pi }{\left( 1+l\right) \Lambda },
\end{equation}%
where our results indicate that there are no positive physical limitation
points for the AdS case (i.e. $\Lambda <0$), because $l>-1$. It means that
the temperature cannot be negative for AdS case when $l>-1$.

To study the phase transition critical points (or divergence points of the
heat capacity ($S_{div}$)), we have to solve the relation $\frac{\partial
^{2}M\left( S\right) }{\partial S^{2}}=0$. In other words, we set the
denominator of the heat capacity (Eq. (\ref{Heat1})) equal to zero (i.e., $1+%
\frac{\left( 1+l\right) \Lambda S}{\pi }=0$), which leads to 
\begin{equation}
S_{div}=\frac{-\pi }{\left( 1+l\right) \Lambda },  \label{rdivHeat}
\end{equation}%
where shows a phase transition critical point for AdS black holes, which
depends on the Lorentz-violating parameter $l$ (or the non-minimum
interaction between the bumblebee field and the Ricci tensor). In other
words, when $l>0$, the phase transition critical point decreases as $l$
increases. Conversely, when $-1<l<0$, the critical point increases as the
magnitude of $l$ increases.

\begin{figure}[tbh]
\centering
\includegraphics[width=0.4\textwidth]{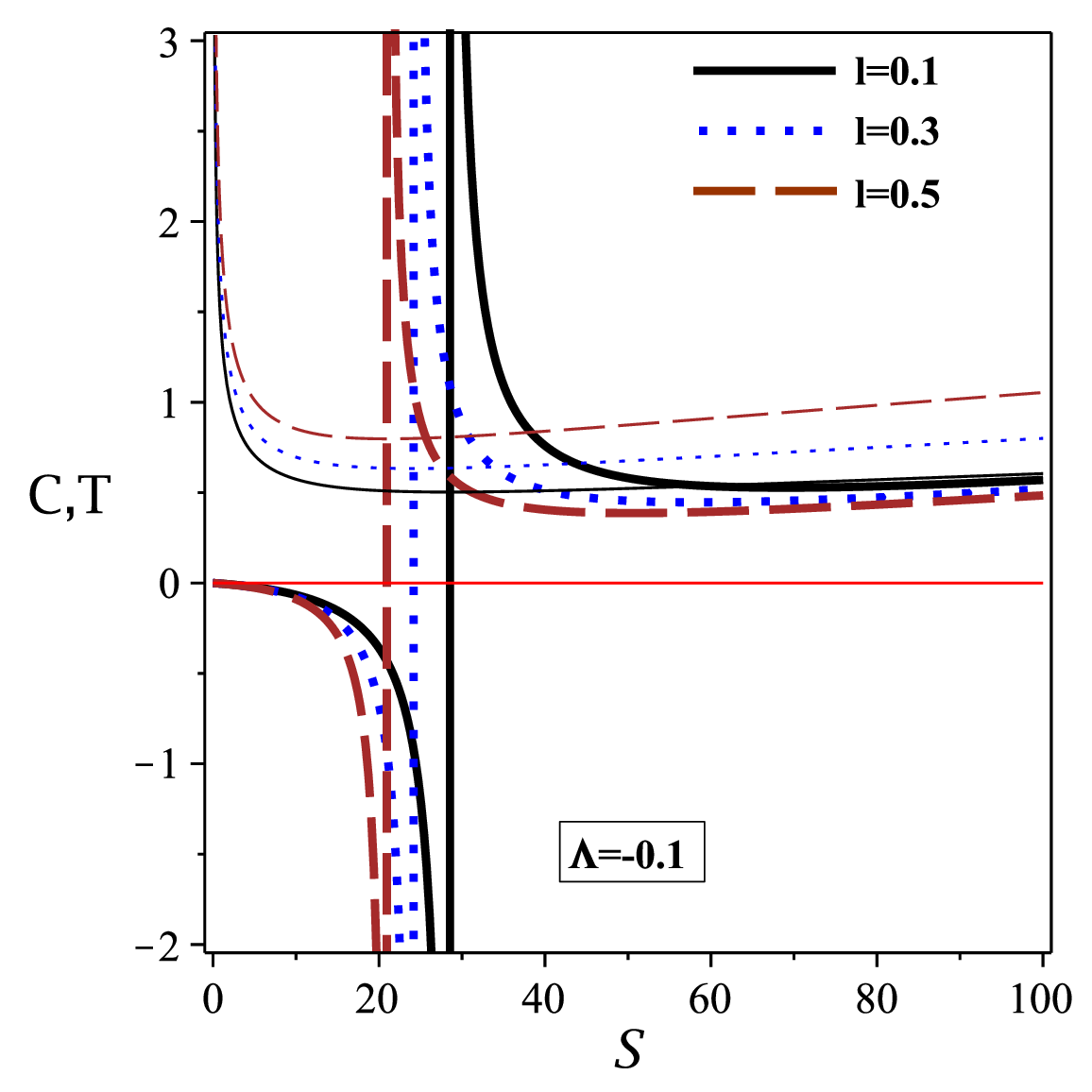} \newline
\includegraphics[width=0.4\textwidth]{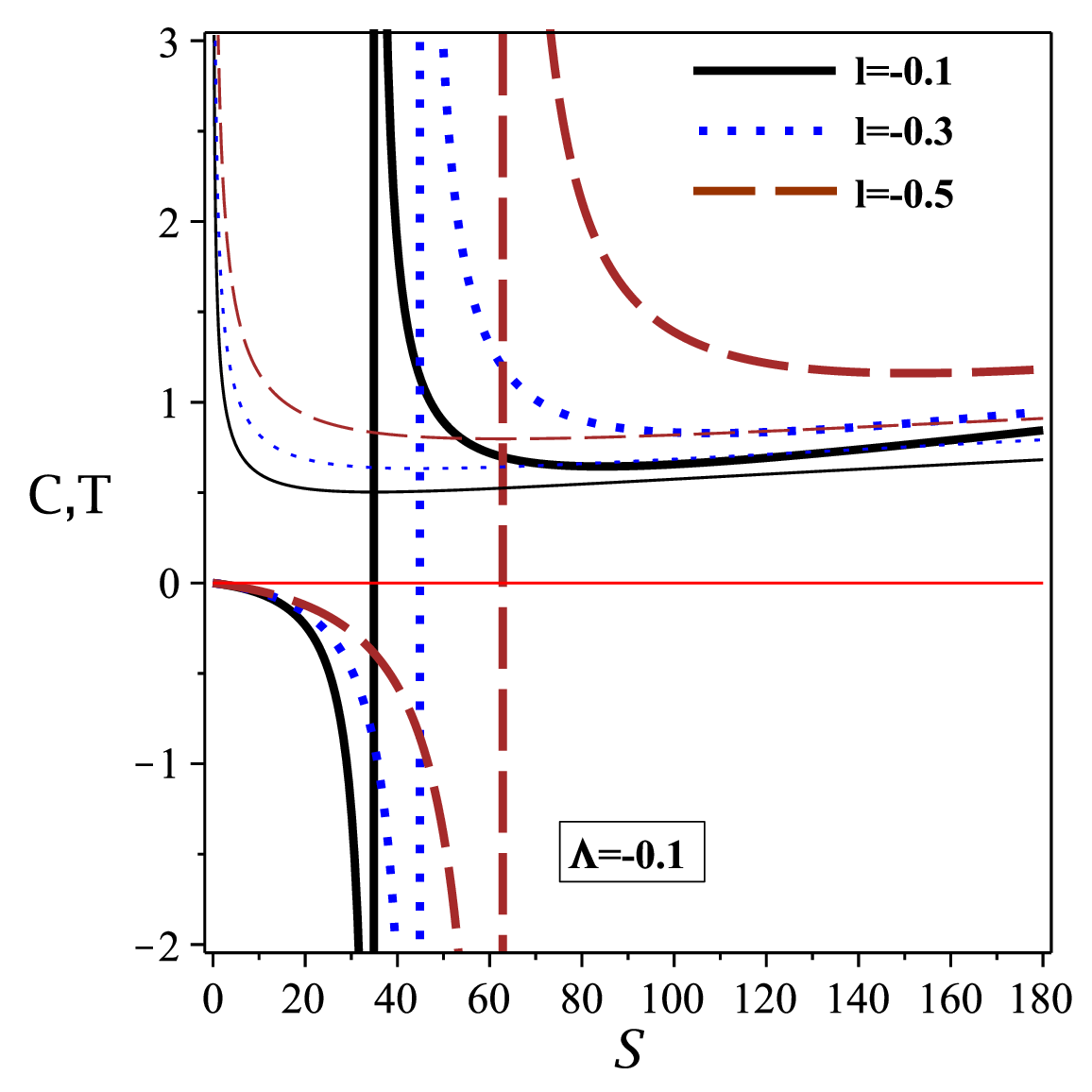} \newline
\caption{$C$ and $T$ versus $S$ for positive value of $l$ (up panel), and
negative value of $l$ (down panel).}
\label{Fig2}
\end{figure}

Now we can assess the local stability by considering the behavior of both
temperature and heat capacity simultaneously. According to our analysis in
Fig. \ref{Fig2}, it is evident that the large black holes meet the criteria
for local stability as both the heat capacity and temperature values are
positive. Additionally, when $l>0$ ($-1<l<0$), the area of local stability
increases (decreases) as $|l|$ increases.

\subsubsection{Global stability}

In the context of the canonical ensemble, the global stability of a
thermodynamic system can be studied by Helmholtz free energy. In other
words, the negative of the Helmholtz free energy determines the global
stability of a thermodynamic system. Therefore, by using the Helmholtz free
energy, we want to evaluate the global stability of the AdS black holes in
bumblebee gravity.

\begin{figure}[tbh]
\centering
\includegraphics[width=0.4\textwidth]{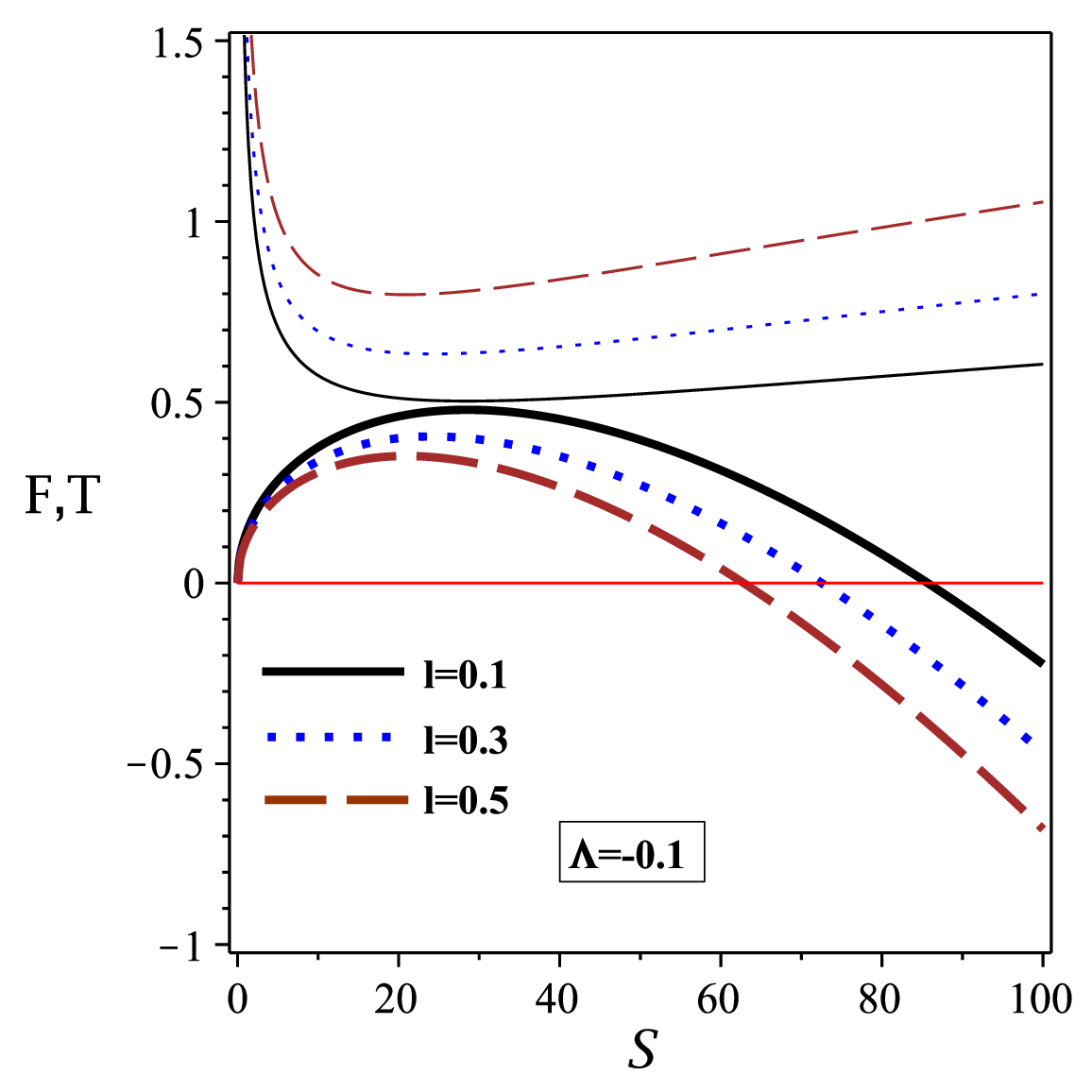} \newline
\includegraphics[width=0.4\textwidth]{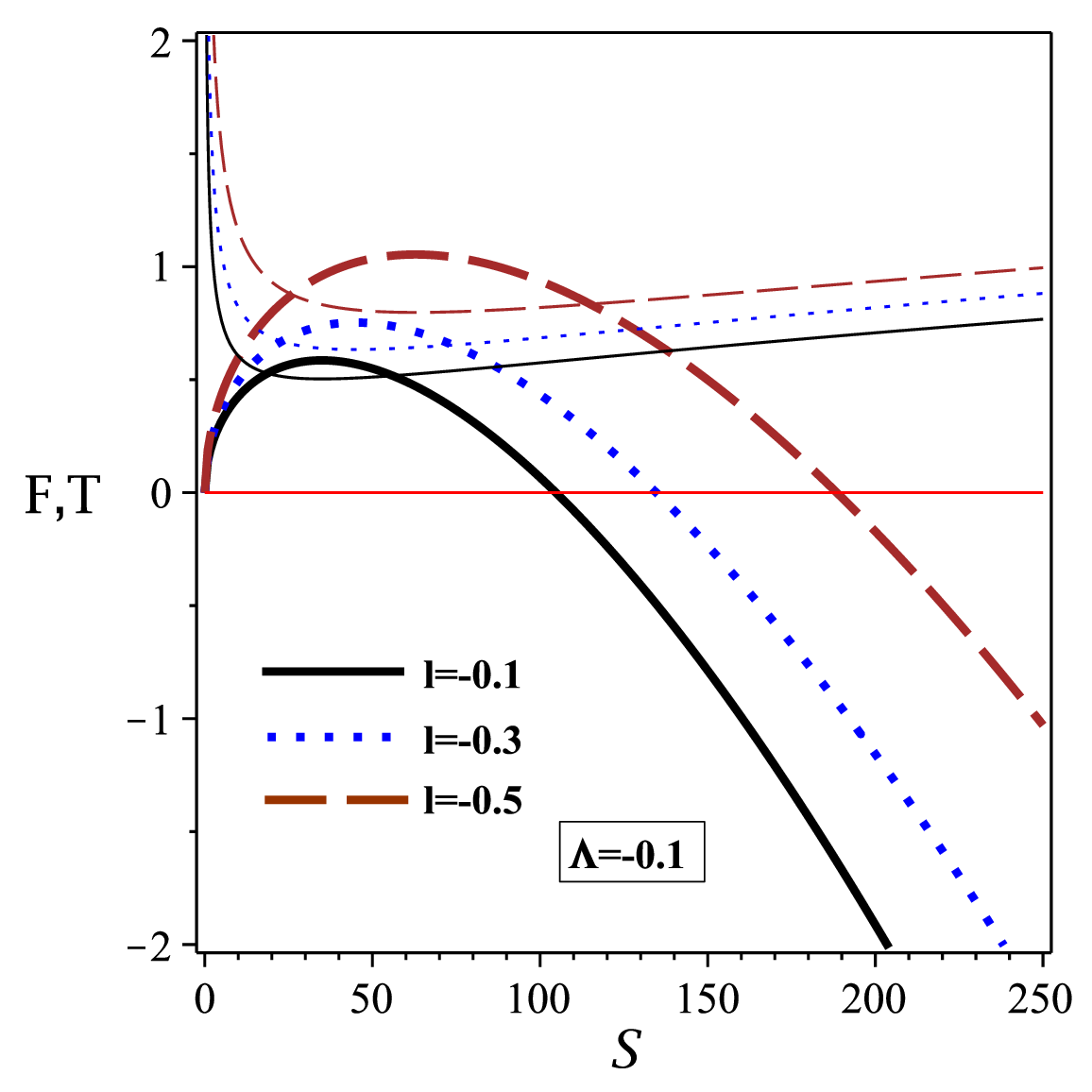} \newline
\caption{$F$ and $T$ versus $S$ for positive value of $l$ (up panel), and
negative value of $l$ (down panel).}
\label{Fig3}
\end{figure}

In the usual case of thermodynamics the Helmholtz free energy is given by $%
F=U-TS$. However, in the context of the black holes, Helmholtz free energy
is defined in the form $F(T,S)=M\left( S\right) -TS$, where by considering
Eqs. (\ref{MSQ}) and (\ref{TM}), we can obtain the Helmholtz free energy as 
\begin{equation}
F(T,S)=\frac{\left( 1+\frac{\left( 1+l\right) \Lambda S}{3\pi }\right) \sqrt{%
S}}{4\pi ^{1/2}\sqrt{1+l}},
\end{equation}%
and by solving $F(T,S)=0$, we get the roots of the Helmholtz free energy
that is 
\begin{equation}
S_{F}=\frac{-3\pi }{\left( 1+l\right) \Lambda },
\end{equation}%
which indicates that there is a real positive root for the Helmholtz free
energy when $\Lambda <0$. Notably, this real positive root also depends on
the bumblebee field. The global stability areas are given when the Helmholtz
free energy is negative (i.e., $F(T,S)<0$). For this purpose, we plot $%
F(T,S) $ versus $S$ in Fig. \ref{Fig3}. Our findings in Fig. \ref{Fig3}
reveal that: i) the large bumblebee AdS black holes satisfy the global
stability similar to local stability, and ii) the global stability area
increases (decreases) by increasing $l$ ($\left\vert l\right\vert $) when $%
l>0$ ($-1<l<0$).

In summary, our research shows that the large bumblebee AdS black holes
possess both local and global stabilities simultaneously. Moreover, the
existence of the Lorentz-violating parameter ($l$) affects the size of the
stable area.

\subsection{Thermal stability in extended phase space}

In extended phase space the effective cosmological constant plays the role
of a thermodynamic pressure, given by 
\begin{equation}
P=\frac{-\Lambda \left( 1+l\right) }{8\pi }.  \label{Lambda}
\end{equation}%
The Lorentz-violating parameter ($l$) is shown to have an effect on the
pressure, as indicated by the equation (\ref{Lambda}). Depending on the
value of the Lorentz-violating parameter, two different behaviors can be
observed for the modified thermodynamic pressure: i) When $l>0$, the
pressure increases as $l$ increases. ii) Conversely, when $-1<l<0$, the
pressure decreases as the magnitude of $l$ increases.

By replacing Eq. (\ref{Lambda}) into Eq. (\ref{MSQ}), the total mass of the
black hole in terms of the entropy and the thermodynamic pressure is%
\begin{equation}
M\left( S,P\right) =\frac{\left( 3+8PS\right) \sqrt{S}}{6\pi ^{1/2}\sqrt{1+l}%
}.  \label{MSP}
\end{equation}

We can get the thermodynamic volume ($V$) of bumblebee AdS black holes by
using Eq. (\ref{MSP}), which leads to 
\begin{equation}
V=\left( \frac{\partial M}{\partial P}\right) _{S}=\frac{4\pi }{3\sqrt{
\left( 1+l\right) }}\left( \frac{S}{\pi }\right) ^{3/2}.  \label{V1}
\end{equation}

Using $M\left( S,P\right) $, we can obtain the temperature ($T$), the heat
capacity at constant pressure ($C_{P}$), the heat capacity at constant
volume ($C_{V}$) in the following forms%
\begin{eqnarray}
T &=&\left. \frac{\partial M}{\partial S}\right\vert _{P}=\frac{1+8PS}{4%
\sqrt{\pi S\left( 1+l\right) }},  \label{Text} \\
&&  \notag \\
C_{P} &=&\frac{T}{\left. \frac{\partial T}{\partial S}\right\vert _{P}}=%
\frac{2S\left( 1+8PS\right) }{8PS-1},  \label{Cpext} \\
&&  \notag \\
C_{V} &=&\left. T\left( \frac{\partial S}{\partial T}\right) \right\vert
_{V}=0.  \label{Cvext}
\end{eqnarray}

The temperature (Eq. (\ref{Text})) impose a constraint on $l$. In other
words, to avoid of imaginary value of the temperature we have to consider $%
l>-1$. In other words, to have physical black holes, the temperature must be
real positive value, so we have to respect to limit $l>-1$. Notably, $C_{V}$
is zero for static black holes in any modified theory of gravity when the
thermodynamic volume depends on entropy, i.e., $V\propto S$ (see Eq. \ref{V1}%
)). Indeed, using the definition of heat capacity at constant volume $%
C_{V}=T $ $\left. \left( \frac{\partial S}{\partial T}\right) \right\vert
_{V}=\frac{\left( \frac{\partial M(S,P)}{\partial S}\right) \bigg|_{V}}{%
\left( \frac{\partial T(S,P)}{\partial S}\right) \bigg|_{V}}$, and
considering that $V\propto S$ (see Eq. (\ref{V1})), we find that $T=\left(%
\frac{\partial M(S,P)}{\partial S}\right) \bigg|_{V}=0$, which leads to $%
C_{V}=0$.

To study the thermodynamic stability of the system in the extended phase
space, we focus on the heat capacity at constant pressure ($C_{P}$). For $%
C_{P}$ to be positive, we need to satisfy the condition $8PS-1 > 0$.
Substituting the relation $P = \frac{-\Lambda(1+l)}{8\pi}$ into this
condition leads to the inequality $l<\frac{\pi }{\Lambda S}-1$. This result
implies that to meet the condition $8PS - 1 > 0$, we must have $l<-1$ when $%
\Lambda<0$, which is inconsistent with the conditions that require $l>-1$
for both temperature and total mass.

\subsection{Super-entropy black holes}

In Ref. \cite{Cong2019, Johnson2020}, it was suggested that there is a link
between super-entropy black holes and thermodynamic instability. Confirming
this connection would be of great importance. It should be mentioned that
super-entropy black holes are a result of the violation of the inverse
isoperimetric inequality \cite{Cvetic2011}. A black hole is considered a
super-entropy black hole if it meets the condition $\mathcal{R}<1$. This
ratio is defined for $4-$dimensional spacetime as \cite{Cvetic2011} 
\begin{equation}
\mathcal{R}=\left( \frac{3V}{4\pi }\right) ^{1/3}\left( \frac{4\pi }{A}%
\right) ^{1/2},  \label{R}
\end{equation}%
where $A=4S$ is the horizon area.

After substituting the thermodynamic volume (\ref{V1}) and the horizon area
into Eq. (\ref{R}), we get 
\begin{equation}
\mathcal{R}=\left( 1+l\right) ^{-1/6},
\end{equation}%
which indicates that $\mathcal{R}$ can be less than $1$ (i.e., $\mathcal{R}%
<1 $) for positive values of $l$. Therefore, the bumblebee AdS black holes
are considered super-entropy black holes when the bumblebee field is
positive. This implies that the bumblebee AdS black holes are thermodynamic
instability systems when $l>0$. This is consistent with our analysis because 
$C_{P}$ cannot be positive when $l>0$. In other words, bumblebee AdS black
holes cannot satisfy the stability conditions in the extended phase space.
Consequently, super-entropy is associated with thermodynamic instability.

\section{\textbf{Conclusions}}

In this paper, we first reviewed the bumblebee AdS black holes in Einstein's
gravity. Then, we evaluated the effect of the Lorentz-violating parameter on
the event horizon. Our analysis in Fig. \ref{Fig1} indicated that for the
positive Lorentz-violating parameter ($l>0$), the radius of the event
horizon decreased as $l$ increased. However, for the negative
Lorentz-violating parameter ($l<0$), the event horizon increased as $|l|$
increased.

We obtained thermodynamic quantities such as the Hawking temperature,
entropy, and total mass for the bumblebee AdS black holes. We found that
both the Hawking temperature and total mass depend on the Lorentz-violating
parameter. To determine the positive values for the Hawking temperature and
total mass, we considered the permissible values of the Lorentz-violating
parameter, which are within the range $l>-1$. Furthermore, we indicated that
these conserved and thermodynamic quantities satisfy the first law of
thermodynamics.

We studied the heat capacity to evaluate local stability. Our results
revealed that there were no physical limiting points for the bumblebee AdS
black holes. In other words, the temperature of bumblebee AdS black holes
was always positive. We obtained the divergences of the heat capacity to
determine the phase transition critical points. Our analysis indicated that
there was a phase transition critical point for AdS black holes that
depended on the Lorentz-violating parameter. In other words, the phase
transition critical point decreased (increased) as $|l|$ increased when $l>0$
($-1<l<0$). Furthermore, the local stability area increased as $l$ increased
when $l>0$.

We evaluated the Helmholtz free energy to investigate global stability. We
found a real positive root for the Helmholtz free energy when $\Lambda <0$,
which also depended on the Lorentz-violating parameter. Our findings in Fig. %
\ref{Fig3} indicated that large bumblebee AdS black holes could satisfy the
global stability condition. Additionally, this global stability area
increased as $l$ increased when $l>0$. Our analysis of the heat capacity and
the Helmholtz free energy simultaneously revealed that large bumblebee AdS
black holes satisfied both local and global stability. Furthermore, these
stable areas depended on the Lorentz-violating parameter.

In the extended phase space, we examined the heat capacity and temperature
to determine the stable regions of bumblebee AdS black holes. The
temperature imposed a constraint on $l$, which was $l>-1$. On the other
hand, to ensure a positive value for the heat capacity at constant pressure (%
$C_{P}$), we found that $l<-1$. Our analysis from these quantities revealed
that these black holes could not satisfy the thermodynamic stability because
the temperature and the heat were not positive, simultaneously. Therefore,
there was no thermodynamic stability for bumblebee AdS black holes.

We found that the bumblebee AdS black holes were super-entropy black holes
when the Lorentz-violating parameter was positive. This revealed the effect
of the Lorentz-violating parameter on the condition of super-entropy black
holes. Additionally, our analysis using $C_{P}$ indicated that the bumblebee
AdS black holes were thermodynamically unstable systems, which is consistent
with the conjecture regarding super-entropy black holes. Therefore, the
bumblebee AdS black hole satisfied the conditions of thermodynamic
instability and was classified as a super-entropy black hole. In fact, we
confirmed that the bumblebee AdS black hole met the conjecture presented by
Cong and Mann \cite{Cong2019}.

\begin{acknowledgements}
B. Eslam Panah thanks University of Mazandaran.
\end{acknowledgements}


\begin{thebibliography}{99}
\bibitem{Maluf2021} R. V. Maluf, and J. C. S. Neves, Phys. Rev. D \textbf{103%
}, 044002 (2021).

\bibitem{Bailey2006} Q. G. Bailey, and V. A. Kostelecky, Phys. Rev. D 
\textbf{74}, 045001 (2006).

\bibitem{Malufetal2013} R. V. Maluf, V. Santos, W. T. Cruz, and C. A. S.
Almeida, Phys. Rev. D \textbf{88}, 025005 (2013).

\bibitem{KosteleckyS1989a} V. A. Kostelecky, and S. Samuel, Phys. Rev. D 
\textbf{39}, 683 (1989).

\bibitem{KosteleckyS1989b} V. A. Kostelecky, and S. Samuel, Phys. Rev. Lett. 
\textbf{63}, 224 (1989).

\bibitem{KosteleckyS1989c} V. A. Kostelecky, and S. Samuel, Phys. Rev. D 
\textbf{40}, 1886 (1989).

\bibitem{KosteleckyP1991} V. A. Kostelecky, and R. Potting, Nucl. Phys. B 
\textbf{359}, 545 (1991).

\bibitem{GambiniP1999} R. Gambini, and J. Pullin, Phys. Rev. D \textbf{59},
124021 (1999).

\bibitem{Ellis2000} J. R. Ellis, N. E. Mavromatos, and D. V. Nanopoulos,
Gen. Relativ. Gravit. \textbf{32}, 127 (2000).

\bibitem{Carroll2001} S. M. Carroll, J. A. Harvey, V. A. Kostelecky, C. D.
Lane, and T. Okamoto, Phys. Rev. Lett. \textbf{87}, 141601 (2001).

\bibitem{Mocioiu2000} I. Mocioiu, M. Pospelov, and R. Roiban, Phys. Lett. B 
\textbf{489}, 390 (2000).

\bibitem{Ferrari2007} A. F. Ferrari, M. Gomes, J. R. Nascimento, E. Passos,
A. Yu. Petrov, and A. J. da Silva, Phys. Lett. B \textbf{652}, 174 (2007).

\bibitem{Horava2009} P. Horava, Phys. Rev. D \textbf{79}, 084008 (2009).

\bibitem{Rizzo2010} T. G. Rizzo, J. High Energy Phys. \textbf{1011}, 156
(2010).

\bibitem{Santos2013} V. Santos, and C. A. S. Almeida, Phys. Lett. B \textbf{%
718}, 1114 (2013).

\bibitem{MagueijoS2004} J. Magueijo, and L. Smolin, Class. Quantum Gravit. 
\textbf{21}, 1725 (2004).

\bibitem{Colladay1997} D. Colladay and V. A. Kostelecky, Phys. Rev. D 
\textbf{55}, 6760 (1997).

\bibitem{Colladay1998} D. Colladay and V. A. Kostelecky, Phys. Rev. D 
\textbf{58}, 116002 (1998).

\bibitem{Bluhm2005} R. Bluhm, and V. A. Kostelecky, Phys. Rev. D \textbf{71}%
, 065008 (2005).

\bibitem{Bluhm2008} R. Bluhm, S. -H. Fung, and V. A. Kostelecky, Phys. Rev.
D \textbf{77}, 065020 (2008).

\bibitem{Bluhm1997} R. Bluhm, V. A. Kostelecky, and N. Russell, Phys. Rev.
Lett. \textbf{79}, 1432 (1997).

\bibitem{Bluhm1999} R. Bluhm, V. A. Kostelecky, and N. Russell, Phys. Rev.
Lett. \textbf{82}, 2254 (1999).

\bibitem{Bluhm2000} R. Bluhm, V. A. Kostelecky, and C. D. Lane, Phys. Rev.
Lett. \textbf{84}, 1098 (2000).

\bibitem{Bluhm2002} R. Bluhm, V. A. Kostelecky, C. D. Lane, and N. Russell,
Phys. Rev. Lett. Bluhm, 090801 (2002).

\bibitem{Carroll1990} S. M. Carroll, G. B. Field, and R. Jackiw, Phys. Rev.
D 41, 1231 (1990).

\bibitem{Andrianov1998} A. A. Andrianov, R. Soldati, and L. Sorbo, Phys.
Rev. D \textbf{59}, 025002 (1998).

\bibitem{Lehnert2004} R. Lehnert and R. Potting, Phys. Rev. Lett. \textbf{93}%
, 110402 (2004).

\bibitem{Kaufhold2006} C. Kaufhold, and F. R. Klinkhamer, Nucl. Phys. B 
\textbf{734}, 1 (2006).

\bibitem{Belich2013} H. Belich, L. D. Bernald, P. Gaete, and J. A.
Helayel-Neto, Eur. Phys. J. C \textbf{73}, 2632 (2013).

\bibitem{Colladay2001} D. Colladay, and V. A. Kostelecky, Phys. Lett. B 
\textbf{511}, 209 (2001).

\bibitem{Lehnert2004b} R. Lehnert, J. Math. Phys. \textbf{45}, 3399 (2004).

\bibitem{Altschul2004} B. Altschul, Phys. Rev. D \textbf{70}, 056005 (2004).

\bibitem{Shore2005} G. M. Shore, Nucl. Phys. B \textbf{717}, 86 (2005).

\bibitem{Jackiw1999} R. Jackiw, and V. A. Kostelecky, Phys. Rev. Lett. 
\textbf{82}, 3572 (1999).

\bibitem{Chung2001} J. M. Chung, and B. K. Chung, Phys. Rev. D \textbf{63},
105015 (2001).

\bibitem{Battistel2001} O. A. Battistel, and G. Dallabona, Nucl. Phys. B 
\textbf{610}, 316 (2001).

\bibitem{Scarpelli2001} A. P. B. Scarpelli, M. Sampaio, M.C. Nemes, and B.
Hiller, Phys. Rev. D \textbf{64}, 046013 (2001).

\bibitem{Mariz2005} T. Mariz, J. R. Nascimento, E. Passos, R. F. Ribeiro,
and F. A. Brito, J. High Energy Phys. \textbf{0510}, 019 (2005).

\bibitem{Nascimento2007} J. R. Nascimento, E. Passos, A. Yu. Petrov, and F.
A. Brito, J. High Energy Phys. \textbf{0706}, 016 (2007).

\bibitem{Scarpelli2008} A. P. B. Scarpelli, M. Sampaio, M. C. Nemes, and B.
Hiller, Eur. Phys. J. C \textbf{56}, 571 (2008).

\bibitem{Cima2010} O. M. Del Cima, J. M. Fonseca, D. H. T. Franco, and O.
Piguet, Phys. Lett. B \textbf{688}, 258 (2010).

\bibitem{Gazzola2012} G. Gazzola, H. G. Fargnoli, A. P. Baeta Scarpelli, M.
Sampaio, and M. C. Nemes, J. Phys. G \textbf{39}, 035002 (2012).

\bibitem{Baeta2012} A. P. Baeta Scarpelli, J. Phys. G \textbf{39}, 125001
(2012).

\bibitem{Agostini2012} B. Agostini, F. A. Barone, F. E. Barone, P. Gaete,
and J. A. Helayel-Neto, Phys. Lett. B \textbf{708}, 212 (2012).

\bibitem{Brito2013} L. C. T. Brito, H. G. Fargnoli, and A. P. Baeta
Scarpelli, Phys. Rev. D \textbf{87}, 125023 (2013).

\bibitem{Ovgun2019} A. Ovgun, K. Jusufi, and I. Sakalli, Phys. Rev. D 
\textbf{99}, 024042 (2019).

\bibitem{Oliveira2019} R. Oliveira, D. M. Dantas, V. Santos, and C. A. S.
Almeida, Class. Quantum Grav. \textbf{36}, 105013 (2019).

\bibitem{Kanzi2019} S. Kanzi, and I. Sakalli, Nucl. Phys. B \textbf{946},
114703 (2019).

\bibitem{Kanzi2021} S. Kanzi, and I. Sakalli, Eur. Phys. J. C \textbf{81},
501 (2021).

\bibitem{Kanzi2022} S. Kanzi, and I. Sakalli, Eur. Phys. J. C \textbf{82},
93 (2022).

\bibitem{Mustafa2025} G. Mustafa, et al., Phys. Dark Universe \textbf{47},
101753 (2025).

\bibitem{Shadow1} M. Afrin, S. G. Ghosh, and A. Wang, Phys. Dark Universe. 
\textbf{46}, 101642 (2024).

\bibitem{Shadow2} S. Ul Islam, S. G. Ghosh, and S. D. Maharaj,
[arXiv:2410.05395].

\bibitem{BHBB1} Z. Li, and A. Ovgun, Phys. Rev. D \textbf{101}, 024040
(2020).

\bibitem{BHBB2} S. Chen, M. Wang, and J. Jing, JHEP \textbf{07}, 054 (2020).

\bibitem{BHBB3} Z. Wang, S. Chen, and J. Jing, Eur. Phys. J. C \textbf{82},
528 (2022).

\bibitem{BHBB4} J. Gu, et al., Eur. Phys. J. C \textbf{82}, 708 (2022).

\bibitem{BHBB5} A. Uniyal, S. Kanzi, and A. Sakalli, Eur. Phys. J. C 83, 668
(2023).

\bibitem{Capelo2015} D. Capelo, and J. Paramos, Phys. Rev. D \textbf{91},
104007 (2015).

\bibitem{MalufJCAP2021} R. V. Maluf, and J. C. S. Neves, JCAP \textbf{10},
038 (2021).

\bibitem{Sarmah2024} P. Sarmah, and U. Dev Goswami, [arXiv:2407.13487].

\bibitem{CosBB1} L. A. Lessa, J. E. G. Silva, and C. A. S. Almeida, EPL 141,
29001 (2023).

\bibitem{CosBB2} J. C. S. Neves, Ann. Phys. \textbf{454}, 169338 (2023).

\bibitem{CosBB3} X. Zhu, R. Xu, and D. Xu, [arXiv:2411.18559].

\bibitem{AMDI} A. Ashtekar, and A. Magnon, Class. Quantum Gravit. \textbf{1}%
, L39 (1984).

\bibitem{AMDII} A. Ashtekar, and S. Das, Class. Quantum Gravit. \textbf{17},
L17 (2000).

\bibitem{EslamPanah2018} B. Eslam Panah, Phys. Lett. B \textbf{787}, 45
(2018).

\bibitem{Cong2019} W. Cong, and R. B. Mann, J. High Energy Phys. \textbf{11}%
, 004 (2019)

\bibitem{Johnson2020} C. V. Johnson, Mod. Phys. Lett. A \textbf{35}, 2050098
(2020).

\bibitem{Cvetic2011} M. Cvetic, G. W. Gibbons, D. Kubiznak and C. N. Pope,
Phys. Rev. D \textbf{84}, 024037 (2011).
\end{thebibliography}
\end{document}